\documentstyle[multicol,aps]{revtex}

\begin{document}
\draft
\title{Full $0\hbar\omega$ shell model calculation of the
 binding energies of the 1f$_{\frac{7}{2}}$ nuclei}

\author{E. Caurier$^1$, G. Mart\'{\i}nez-Pinedo$^{2,3}$, F.  Nowacki$^4$,
  A. Poves$^5$, J. Retamosa$^6$ and A. P. Zuker$^1$}

\address{$^1$Institut de Recherches
  Subatomiques (IN2P3-CNRS-Universit\'e Louis Pasteur) B\^at.27/1,
  F--67037 Strasbourg Cedex~2, France}

\address{$^2$W.\,K. Kellogg Radiation Laboratory, California Institute
  of Technology, Pasadena CA 91125, USA}

\address{$^3$Institute of Physics and Astronomy, University of {\AA}rhus, DK-8000 {\AA}rhus, Denmark}

\address{$^4$Laboratoire de Physique Th\'eorique de Strasbourg, 3-5
  rue de L'Universit\'e, F--67084 Strasbourg Cedex, France}

\address{$^5$Departamento de F\'{\i}sica Te\'orica C-XI, Universidad
  Aut\'onoma de Madrid, E--28049 Madrid, Spain}

\address{$^6$Departamento de F\'{\i}sica At\'omica y Nuclear,
      Universidad Complutense de Madrid, 
      E-28040 Madrid, Spain.\\}

\date{\today}
\maketitle
\begin{abstract}
  Binding energies and other global properties of nuclei in the middle
  of the $pf$ shell, such as M1, E2 and Gamow-Teller sum rules, have
  been obtained using a new Shell Model code (NATHAN) written in
  quasi-spin formalism and using a $j$-$j$-coupled basis. An extensive
  comparison is made with the recently available Shell Model Monte
  Carlo results using the effective interaction KB3. The binding
  energies for -nearly- all the 1f$_{7/2}$ nuclei are compared with the
  measured (and extrapolated) results.
\end{abstract}
\pacs{PACS number(s):21.10.Dr,21.60.Cs}

\begin{multicols}{2}

\section{introduction}
\label{sec:intro}

Detailed Shell Model calculations in the full $pf$-shell have been
recently carried out~\cite{a48,cr48,a4749} up to $A= 52$ using a 
realistic G-matrix \cite{kb} with the monopole modifications of ref.
\cite{pozu} (KB3). These calculations could only be done due to the
availability of the $m$-scheme code ANTOINE \cite{antoine}.
It has allowed us to tackle the largest dimensionalities ever reached by
any exact diagonalization shell model code~\cite{fe52}.  The main
disadvantage of ANTOINE is that $J$ and $T$ are not good quantum
numbers and the dimensions of the matrices are maximal.

One is then led to develop new tools to deal with the increasingly
large model spaces needed in shell-model calculations. In this paper
we present the first results obtained using a new code
[NATHAN]~\cite{coup1,coup2}, that works in $j$-$j$ coupling scheme and
uses the quasi-spin formalism. This code retains the main idea of the
code ANTOINE to calculate efficiently all non-zero matrix elements
during the diagonalization procedure.  It can be used either for
unrestricted calculations as it is the case here, or
for nuclei in which seniority truncations are physically sound as in
the Sn region. The use of a $j$-$j$ coupling scheme allows us to
reduce the the memory requirements with the penalty of an increase in
CPU time. This increase is not so important as new computers double
his speed every year and new shared-memory parallel machines are now
available that allow for a relatively easy parallelization.

This paper has several goals: a) to report on results of the
very large shell model calculations that can now be performed, b) to
use them as benchmarks for the new approximate methods of solving the
large scale shell model problem ---e.g. Monte Carlo Shell
Model~\cite{smmc} or Quantum Monte Carlo diagonalization
method~\cite{qmc}---  c) to analyze the systematics of binding energies
for nuclei $40 \leq A \leq 56$, $20 \leq N,Z \leq 28$, extending
 the already published
results for $A=47$, 48 and 49 down to the beginning of the shell and
up to the N=Z=28 closure, studying the effect of the scaling with the
mass of the matrix elements.  These three objectives will
be dealt with in sections II, III and IV.

\section{The Shell Model Code 'NATHAN' }
\label{sec:code}

For a long time, shell model calculations have been limited to ligth
nuclei or to heavier ones with only a few particles outside closed
shells. Besides the well known problems related to the determination
of a good effective interaction for large valence spaces, there are
compelling technical limitations due to the explosive increase of the
dimensions of the matrices to diagonalize.  The diagonalization in
itself is not a problem since, in general, only a few eigenvectors are
needed and in this case the Lanczos method is very efficient.  For
very large matrices, the convergence of the method is optimized by
preliminary calculations in a truncated space.  The fundamental
problem is that we have to deal with ``giant'' matrices, giant meaning
that the number of its non-zero elements is so large that it is
impossible to store all of them before doing the Lanczos procedure.
For this reason, one needs to compute all the non-zero terms at each
new Lanczos iteration.  It is clear that modern shell model codes must
tackle this problem and that the quality of the code will be directly
related to its performance in the calculation of non-zero terms during
the Lanczos procedure itself.

The first breakthrough in this direction was due to the Glasgow
group~\cite{white}.  They took advantage of the simplicity of the
m-scheme. In their code, each Slater determinant is represented by an
integer word and each individual state by a bit in this word. Bit
manipulation and bisection algorithms allow for a fast generation of
the matrix elements. The shell model code ANTOINE adds some important
improvements to the Glasgow method.  The basic idea is to separate the
total space in a product of the smaller spaces spanned by protons and
neutrons.  Then a state $I$ in the global basis can be labelled by a
pair of proton($i$) and neutron ($\alpha$) states. All the
$i$($\alpha$) states are classified in blocks defined by their $J_z$
value.  To any proton block $J_z^p$ corresponds a neutron block
$J_z^n=M-J_z^p$, where $M$ is the total angular momentum projection.
Total wave functions are built by the association of  a proton state $i$
(belonging to the block $J_z^p$) to a neutron state $\alpha$
(belonging to the corresponding block $J_z^n=M-J_z^p$).
 A simple numerical relation
\begin{equation}
 I=R(i)+\alpha
\end{equation}
can be established.  Non-zero terms of the matrix
  are obtained through 3 integer additions:
$I=R(i)+\alpha$, $J=R(j)+\beta$ and $H_{IJ}=V(K)$ with 
$K=Q(q)+\mu$. $q(\mu)$indices the one-body
operator acting between $i(\alpha)$ and $j(\beta)$ states.
($i$,$j$,$q$) and ($\alpha$,$\beta$,$\mu$) are precalculated
with the Glasgow method. The storage of these labels remains possible as the
dimensions in respective proton and neutron spaces are moderated
compared to those of the full space.

Another improvement that the code ANTOINE incorporates is an initial
Lanczos procedure with the operators $J^2$ and $T^2$, i.e. a
projection onto good $J$ and $T$.  Basis states of good $J$ and $T$ are
then used as initial states for the hamiltonian's Lanczos iterations.
This accelerates the convergency dramatically.

The main disadvantage of the m-scheme is that the space comprises all
the states with $J \ge J_z$ and $T \ge T_z$. The fundamental 
limitation is the capacity of storage the Lanczos vectors. For that 
reason, we have thought to adapt the idea of separating proton and
neutron spaces to a coupled basis. Now, each $i$ and $\alpha$
functions are coupled to good angular momentum with the usual
techniques of the Oak-Ridge/Rochester code \cite{french}.
The  $i$ and $\alpha$ states are classified by  their $J$
values.  The fundamental difference with respect to the m-scheme is that now
\begin{itemize}
\item to a
$J_p$ value correspond all the blocks $J_n$ with $|J_n-J_p| \leq J \leq
J_n+J_p$.However, the relation $I=R(i)+\alpha$ remains valid.
\item  The N-body matrix elements are now given by the relation $H_{IJ}=h_{ij}.h_{\alpha\beta}.W(K)$, 
$h_{ij}(h_{\alpha\beta})$ being one-body matrix elements
calculated with the Quasi-spin formalism and $W(K)$ 
being the product of a two-body matrix element with some 
$9j$ recoupling coeeficient.
We notice that to the 3 previous additions to generate $H_{IJ}$
in the m-scheme case, we have to perform now 3 supplementary products.
\end{itemize}
To summarize, we can say that the two codes are complementary. The coupled
formalism is  more efficient in the following cases:
\begin{itemize}
\item For $J=0^+$ states (the dimension is two orders of magnitude smaller
  than in the m-scheme) and to a lesser extent for low spin states.
\item when we need a lot of Lanczos iterations(for the calculation of strength functions).
\item When seniority truncations are reasonable.
\item When the size of the m-scheme Lanczos vectors exceeds
  the storage capacity of the disks.
\end{itemize}

In other cases, the m-scheme code ANTOINE remains a 
better option.  The two codes run on ordinary workstations. Indeed, the
use of parallel computers should improve strongly their performances.

The  code  NATHAN  has made 
it possible to carry out calculations that, if made in $m$-scheme,
would involve more than one billion M=0 Slater determinants, as in our
calculation of the ground state of $^{56}$Ni in the full $pf$-shell.
The dimensions of the J=0 matrices together with their equivalent
$m$-scheme dimensions are listed in table I for some of the nuclei we
have studied in this work. Once the energy and the wave function of
the ground state of a given nucleus is obtained, it is easy
to built the doorway states (also named sum rule states) acting with
the different transition operators $\Omega^{\lambda}$ on it. The norm
of the doorway gives  the non-energy weighted sum rule for the
operator. If the doorway is used as starting vector in the Lanczos
process, successive iterations provide the energy to the n-th weighted
sum rules or equivalently the different moments of the strength
function of the transition operator chosen. Notice that already with
two iterations we  have the norm, the centroid, the width and the
skewness of the distribution of strength. These are averaged
quantities that can also be accessed by the new stochastic approaches
to the Shell Model problem, as for instance the Shell Model Monte
Carlo (SMMC), and we shall devote the next section to compare the
approximate and exact solutions.

\section{Benchmarks and comparison with SMMC results}

With the advent of the stochastic approximations to the solution of
the Shell Model problem, mentioned above, it becomes compulsory to
dispose of large enough sets of exacts results in order to benchmark
the accuracy of the new methods and to uncover their strong and weak
aspects. We have chosen to make this comparison with the set of nuclei
studied by Langanke {\it et al.}~\cite{langan} using Caltech's SMMC.
The effective interaction KB3 is used troughout, with effective
charges 1.35 for protons and 0.35 for neutrons, bare $g$-factors and
unquenched Gamow-Teller operator. The choice of an isoscalar effective
charge of 1.7 in \cite{langan} instead of the canonical value of 2,
leads to values that underpredict the experimental quadrupole
transition rates. However this is irrelevant for our purpose of
comparing SMMC and exact SM diagonalizations. SMMC involves two
extrapolations; one in temperature and another in the parameter that
has to be introduced \cite{alhasi} in order to change the sign of
those terms of the hamiltonian that have ``bad''-sign and that, if
taken at their original value, will spoil the convergence of the Monte
Carlo method. Both extrapolations will contribute to the final
differences with the exact results. While the defaults associated to
the impossibility of doing a zero temperature calculation are smooth
and predictable, those associated to the change of sign of the
``bad''-sign terms are less well under control.

  In table II we gather the energies and the E2 sum rules ($\sum_i
  B(E2) \; 0^+ \rightarrow 2^+_i$). SMMC gives energies that are above the
  exact values by  about 0.5 MeV in most of the cases. This is
  consistent with a residual ``heating'' in SMMC. However, for the
  heaviest part of the set of nuclei studied, the discrepancies grow
  up to reach 2 MeV, indicating problems in the extrapolation linked
  to the ``bad''-sign terms. The E2 sum rules are nicely reproduced by
  SMMC except in a couple of cases, $^{62}$Ni and  $^{64}$Ni where the
  the exact number are clearly missed.

  In table III the comparison is extended to M1 and Gamow-Teller sum
  rules. In most cases, the 10\%--15\% error bars of the SMMC numbers
  suffice to embrace the exact result. Nevertheless, there still
  remain some large deviations in the Gamow-Teller strength of
  $^{60}$Fe, $^{62}$Ni, $^{64}$Ni and $^{64}$Zn.

  The outcome of this comparison is two-sided. On the one side, it
  validates SMMC to the 1--2 MeV level for the ground state energies
  and to the 20\% level for the sum rules. On the other side, there
  are cases in which the discrepacies grow larger without an evident
  cause. This is  a serious thread to the predictive power of SMMC,
  although it is envisageable that a more thorough control of the
  different extrapolations could bring these isolated cases to the
  general pace.

\section{Binding Energies}
\label{sec:bind}

The code NATHAN has given us the opportunity to complete our stock of
binding energies of $pf$-shell nuclei, in the full space, using the
effective interaction KB3. It is our aim now to verify that we can
describe the ground state's energies at the same level of accuracy
that we have achieved for the excitation energies ($\sim$ 200 keV). A
remark is timely here; the monopole part of the interaction KB3 was
fixed only by 1f7/2 nuclei, in a moment when only extremely truncated
calculations were feasible. Therefore its non-1f7/2 monopoles are not
well determined. Furthermore, its quasiparticle gap around $^{56}$Ni
is too strong by about 1 MeV what will results in a relative
underbinding of the nuclei with N or Z larger than 28. That is why
in this section we shall only deal with 1f7/2 nuclei.

What the shell model calculation produces --E(SM) in the second
column of table IV-- is the contribution to the
nuclear binding energy of the interaction of the valence particles
among themselves. It
does not  include the Coulomb repulsion among
the protons, nor the binding energy of the core ($^{40}$Ca in our
case), nor the interaction among the core and the valence particles.
Therefore, in order to compare with the experimental binding
energies relative to $^{40}$Ca, $B_e$, we have  take into account
these quantities.

The Coulomb energies relative to $^{40}$Ca  can be 
calculated using a local formula for a major shell
 ($\pi$ = valence protons, $\nu$ = valence neutrons):

\begin{equation}
  \label{eq:coulomb}
  E_C = e_\pi \pi + V_{\pi\pi} \frac{\pi(\pi-1)}{2} + V_{\pi\nu} \pi\nu.
\end{equation}

In our previous works~\cite{a48,a4749} the values of the constants
$e_\pi$, $V_{\pi\pi}$ and $V_{\pi\pi}$ were determined from the
differences in binding energies between $^{41}$Sc and $^{40}$Ca
($e_\pi$) and the $A=42$ isobars ($V_{\pi\pi}$ and $V_{\pi\nu}$). In
this paper we are interested in larger mass region, hence the need of
a better determination of the constants in
expression~(\ref{eq:coulomb}). In order to do so we have fitted the
Coulomb displacement energies of analog states for nuclei between
$A=42$ and $A=64$~\cite{cde1,cde2}. The resulting parameters are:

\begin{eqnarray}
  \label{eq:parameters}
  e_\pi & = & 7.44 \pm 0.02,\nonumber\\
  V_{\pi\pi} & =& 0.274 \pm 0.003,\\ \nonumber
  V_{\pi\nu} &=& -0.049 \pm 0.003.
\end{eqnarray}

 Another option is to rely in global expressions that are used for the
 Coulomb term of the mass formulas. We have chosen the one used in
 ref~\cite{dufzu};

\begin{eqnarray}
  \label{eq:coulomb1}
  E_C = 0.700 (Z (Z-1) - 0.76 (Z (Z-1))^{2/3})/R_C ;\nonumber\\
  R_C = e^{\frac{1.5}{A}} \cdot A^{\frac{1}{3}} \cdot
             \left(0.946 - 0.573 \cdot \left( \frac{2T}{A}\right)^2\right)
\end{eqnarray}

 The valence space Coulomb energies obtained from~(\ref{eq:coulomb1})
 are very close to those from the local fit (2), with discrepancies
 that never reach 1\%. In what follows we shall use the Coulomb
 energies from the global formula (4).

 Besides, one should add the nuclear interaction between a particle in
 the valence space and the core. The value of this --one body-- matrix
 element is usually taken from the binding energy difference between 
 $^{41}$Ca and $^{40}$Ca. However we shall proceed otherwise; as the
 effective interaction we have been using (KB3), has been only tested
 against spectroscopic observables that will not vary if we add to the
 hamiltonian terms that only depend on scalars made with the {\bf
 total} number of valence particles (n) or the  {\bf total} isospin (T), we
 have the freedom to add the following monopole expression to our
 hamiltonian:

 \begin{equation}
  \label{eq:mono}
  E_M = e_v  \; n + a \; \displaystyle{\frac{1}{2}} n (n-1)  + 
   b \; (T (T+1) - \displaystyle{\frac{3}{4}} \;n)
\end{equation}
   
\noindent
where e$_v$ is an average particle core interaction (hopefully close
to the one experimentally determined in A=41) and a and b are the
isoscalar and isovector global monopole corrections to KB3 that we
will fix by a fit to the experimental binding energies relative to
$^{40}$Ca using the formula:

\begin{equation}
  \label{eq:defbind}
  E_B = - B_e = E(SM)+E_C+E_M,
\end{equation}

The data set is listed in the fourth column of table IV (the
numbers with a star are extrapolated values from \cite{audi} not
included in the fit) and
contains 51 entries. The values of the parameters resulting from the
fit are:

\begin{eqnarray}
  \label{eq:parmono1}
  e_v & = & -8.67  \pm 0.01 \; MeV, \nonumber \\
  a & = &  0.092 \pm 0.003  \; MeV,  \nonumber \\
  b & = &  0.063 \pm 0.006  \; MeV.  \nonumber
\end{eqnarray}

 The Shell Model binding energies calculated with these values are
 listed in the third column of table IV. The
 rms deviation between theory and experiment is 227 keV. These results
 deserve some comments:

\begin{quote}

 -- The rms deviation we have attained fulfils our expectations; we are able
 to describe  consistently at the same level of accuracy excitation
 energies and --valence space-- ground state energies.

 -- The value $e_v$=-8.67 MeV is close enough to the A=41 value -8.36
 MeV as to be considered satisfactory.

 -- The values of the a and b parameters are small and indeed smaller
 than the monopole modifications of some terms of the original
 Kuo-Brown interaction that led to KB3 (about 300 keV).

 -- The Shell Model binding energies for those nuclei not included in 
 the fit are our barest predictions.
  If we compare them with the extrapolated
 values in \cite{audi} the  discrepancies are somewhat larger than for
 the measured values, without exceeding 500~keV in any case.
\end{quote}
  
 There are basic reasons to scale the matrix elements of the
 effective interactions with a term that reflects somehow the change
 in size of the  underlying mean field. In the harmonic oscillator
 basis this brings in the usual A$^{1/3}$ dependence of $\hbar \omega$,
 that has been sometimes incorporated to $sd$ and $pf$-shell effective
 interactions \cite{wilde,richter,retamo}. A more elaborated
 dependence has been proposed recently \cite{zuduf} in order to
 improve the description of nuclear radii. It  leads to the following
 scaling factor:

\begin{equation}
  \label{eq:scale}
  \displaystyle{\left(\frac{A_0}{A}\right)^{\frac{1}{3}}} \cdot
  \displaystyle{e^{3\left(\frac{A-A_0}{A\cdot A_0}\right)}} \cdot
  \displaystyle{\left(\frac{0.946-0.573\left(\frac{2T}{A_0}\right)^2}
                 {0.946-0.573\left(\frac{2T}{A}\right)^2}\right)^{2}}
\end{equation}

\noindent
where A$_0$ is the mass at which the effective interaction has been
computed and A and T are  the mass and the isospin 
of the nucleus we are dealing with.

 It is worth noticing that the lower
$pf$-shell might be special in which has to do with global scalings,
 because the radii of $^{40}$Ca and
$^{58}$Ni can be reproduced without any change in the harmonic
oscillator size parameter and so do the Coulomb displacement energies
\cite{gomez}. On the other side we wondered whether the extra global
monopole correction that comes out of our fit is an artefact due
precisely to the absence of mass dependence in the matrix elements or
not. In order to settle this point we have repeated all the binding
energy calculations with matrix elements scaled as in
equation~(\ref{eq:scale}) with A$_0$=42 (see table V). Afterwords, we
follow exactly the same steps discussed above; we add the same Coulomb
energies and proceed to fit the coefficients of the global monopole
formula~(\ref{eq:mono}), but now with the a and b parameters scaling as
the matrix elements. The values of the parameters at A=42 are:

\begin{eqnarray}
  \label{eq:parmono2}
  e_v & = & -8.61  \pm 0.01 \; MeV, \nonumber \\
  a & = &  0.041 \pm 0.003  \; MeV,  \nonumber \\
  b & = &  0.119 \pm 0.006  \; MeV.  \nonumber
\end{eqnarray}

The resulting binding energies are compared in table V with the
experimental data. The rms deviation is now 215~keV. Therefore we are led
to conclude that the average quality of the agreement is insensitive
to the inclusion of a mass dependence in the two body matrix elements.
Notice that the value of e$_v$ is essentially the same we had 
without mass dependence. On the other hand the a and b parameters are
quite different from the ones we had, even if they are in the same
range of values. It appears that one half of the global isoscalar monopole
correction can be absorbed into the mass dependence, on the contrary
the isovector correction doubles. The predictions for
the binding energies not included in the fit differ from those of the
previous  one typically by 150 keV, in the direction of increasing the
discrepancy with the extrapolated values. Nevertheles none of this
elements is decisive in making a choice between the two approaches. On
the one side, Occam's razor favours the mass independent choice, on
the other side, if we want to go beyond $^{56}$Ni we should surely
need to incorporate the  mass dependence.

\section{Conclusions}

 The new shell model code NATHAN has been used to calculate the
 binding energies, M1, E2 and GT sum rules of several nuclei nuclei of
 the $pf$-shell, in the full valence space, using the effective
 interction KB3. These results have been used to benchmark the SMMC
 calculations, that agree with the exact results within 20\% in most
 cases. We have also computed the binding energy of nearly all
 1f$_{7/2}$ nuclei,
 reaching the same level of agreement than we had for the excitation
 energies and making predictions for a number of still unavailable
 masses. We also show that the inclusion of a mass dependence in the
 two body matrix
 elements is not critical for the description of the binding
 energies in this region.

\bigskip
\noindent
{\bf Acknowledgements}. This work has been partially supported by the
IN2P3-France, CICYT-Spain agreements, and by a DGES (Spain) grant PB96-053.

\narrowtext

\begin{table}[h]
  \begin{center}
    \leavevmode
    \caption{$m$ scheme and $J=0$ dimensions in the full $pf$ shell}
    \label{tab:dimen}
    \begin{tabular}{crr}
      Nucleus & $m$ scheme & $J=0$ dimension \\
      \hline
      $^{48}$Ti & 634\,744   & 14\,177  \\
      $^{50}$Ti & 1\,967\,848   & 39\,899  \\
      $^{52}$Ti & 2\,843\,770   & 55\,944  \\
      $^{50}$Cr & 14\,625\,240  & 267\,054 \\
      $^{52}$Cr & 45\,734\,928  & 773\,549 \\
      $^{54}$Cr & 66\,262\,352  & 1\,093\,850 \\
      $^{52}$Fe & 109\,954\,620 & 1\,777\,116 \\
      $^{54}$Fe & 345\,400\,174 & 5\,220\,621 \\
      $^{56}$Fe & 501\,113\,392 & 7\,413\,488 \\
      $^{56}$Ni & 1\,087\,455\,228 & 15\,443\,684 \\
    \end{tabular}
  \end{center}
\end{table}

\begin{table}[h]
  \begin{center}
    \leavevmode
    \caption{Valence energies and B(E2) sum rules, exact
      diagonalization vs. Shell Model Monte Carlo results}
    \label{tab:be_e2}
    \begin{tabular}{ccccc}
      nucleus &  $E(SM)$ & $E(SM)$ & $\sum B(E2)$ & $\sum B(E2)$   \\
      & (shell model) & (SMMC)& (shell model)&(SMMC)\\ \hline
      $^{48}$Ti & $-$24.6 & $-$23.9  & 476 & 455 $\pm$ 25\\ 
      $^{50}$Ti & $-$27.7 & $-$27.2  & 405 & 465 $\pm$ 50\\ 
      $^{52}$Ti & $-$25.4 & $-$24.9  & 477 & 465 $\pm$ 55\\ 
      $^{54}$Ti & $-$22.0 & $-$21.4  & 445 & 450 $\pm$ 80\\ 
      $^{48}$Cr & $-$32.9 & $-$32.3  & 978 & 945 $\pm$ 45\\ 
      $^{50}$Cr & $-$40.5 & $-$40.0  & 913 & 890 $\pm$ 90\\ 
      $^{52}$Cr & $-$46.0 & $-$45.6  & 690 & 645 $\pm$ 75\\ 
      $^{54}$Cr & $-$47.0 & $-$46.3  & 888 & 890 $\pm$ 90\\ 
      $^{56}$Cr & $-$45.5 & $-$44.8  & 825 & 840 $\pm$ 90\\ 
      $^{52}$Fe & $-$54.3 & $-$53.7  &1016 &1055 $\pm$ 50\\ 
      $^{54}$Fe & $-$62.8 & $-$62.7  & 764 & 750 $\pm$ 80\\ 
      $^{56}$Fe & $-$66.4 & $-$65.8  &1019 & 990 $\pm$ 6\\ 
      $^{58}$Fe & $-$67.7 & $-$66.7  &1117 &1010 $\pm$ 65\\ 
      $^{60}$Fe & $-$67.0 & $-$65.8  &1052 &1105 $\pm$ 65\\ 
      $^{56}$Ni & $-$78.5 & $-$77.8  & 572 & 515 $\pm$ 65\\ 
      $^{62}$Ni & $-$89.5 & $-$87.6  & 823 &1010 $\pm$ 25\\ 
      $^{64}$Ni & $-$89.9 & $-$87.7  & 773 &1165 $\pm$ 80\\ 
      $^{64}$Zn & $-$106.3 & $-$104.8  & 1157 &1225 $\pm$ 65 \\ 
    \end{tabular}
  \end{center}
\end{table}

\begin{table}[h]
  \begin{center}
    \leavevmode
    \caption{
1 and Gamow-Teller sum rules, exact
      diagonalization vs. Shell Model Monte Carlo results}
    \label{tab:bm1_bgt}
    \begin{tabular}{ccccc}
      Nucleus &  $\sum B(M1)$ & $\sum B(M1)$ & $\sum B(GT_+)$  & 
      $\sum B(GT_+)$  \\
      &  (shell model) & (SMMC)& (shell model) & (SMMC)\\
      \hline
      $^{48}$Ti & 10.6 & 10.2 $\pm$ 1.2 & 1.26 & 1.13 $\pm$ 0.18 \\
      $^{50}$Ti & 12.6 & 12.5 $\pm$ 1.0 & 1.24 & 1.47 $\pm$ 0.16 \\
      $^{52}$Ti & 12.9 & 12.5 $\pm$ 1.0 & 0.99 & 1.11 $\pm$ 0.16 \\
      $^{54}$Ti & 13.4 & 13.5 $\pm$ 1.5 & 0.89 & 0.97 $\pm$ 0.21 \\
      $^{48}$Cr & 12.0 & 13.8 $\pm$ 1.7 & 4.13 & 4.37 $\pm$ 0.35 \\
      $^{50}$Cr & 13.9 & 14.5 $\pm$ 2.5 & 3.57 & 3.51 $\pm$ 0.27 \\
      $^{52}$Cr & 15.6 & 18.9 $\pm$ 2.2 & 3.33 & 3.51 $\pm$ 0.19 \\
      $^{54}$Cr & 16.5 & 13.0 $\pm$ 2.5 & 2.24 & 2.21 $\pm$ 0.22 \\
      $^{56}$Cr & 16.3 & 16.2 $\pm$ 2.0 & 1.92 & 1.50 $\pm$ 0.21 \\
      $^{52}$Fe & 17.2 & 18.9 $\pm$ 1.4 & 6.92 & 7.10 $\pm$ 0.42 \\
      $^{54}$Fe & 18.9 & 16.5 $\pm$ 2.8 & 6.33 & 6.05 $\pm$ 0.45 \\
      $^{56}$Fe & 19.4 & 20.4 $\pm$ 3.0 & 4.69 & 3.99 $\pm$ 0.27 \\
      $^{58}$Fe & 18.8 & 20.3 $\pm$ 3.0 & 3.12 & 3.06 $\pm$ 0.28 \\
      $^{60}$Fe & 18.2 & 17.3 $\pm$ 2.1 & 2.60 & 1.80 $\pm$ 0.24 \\
      $^{56}$Ni & 22.8 & 23.0 $\pm$ 1.2 & 10.2 & 9.86 $\pm$ 0.38 \\
      $^{62}$Ni & 20.7 & 19.6 $\pm$ 2.9 & 4.38 & 3.43 $\pm$ 0.40 \\
      $^{64}$Ni & 19.3 & 18.9 $\pm$ 2.7 & 3.44 & 1.73 $\pm$ 0.29 \\
      $^{64}$Zn & 21.6 & 23.6 $\pm$ 2.2 & 5.54 & 4.13 $\pm$ 0.34 \\
    \end{tabular}
  \end{center}
\end{table}

\begin{table}[h]
  \begin{center}
    \leavevmode
    \caption{Shell model binding energies relative to $^{40}$Ca
      compared with experiment. KB3 interaction without
      mass dependence, see text for the details}
    \begin{tabular}{ldddd}
      Nucleus & $E(SM)$ & $B_e$(th) & $B_e$(exp) & $\Delta$ \\
      \hline
      $^{42}$Ca &  $-$2.71 &  19.93 &  19.84  &   0.08 \\
      $^{42}$Sc$^{T=1}$ &  $-$2.71 &  12.44 &  12.64  &  $-$0.20 \\
      $^{42}$Sc$^{T=0}$ &  $-$2.35 &  12.20 &  12.02  &   0.17 \\
      $^{42}$Ti &  $-$2.71 &   4.55 &   4.85  &  $-$0.30 \\
      $^{43}$Ca &  $-$2.55 &  28.19 &  27.78  &   0.41 \\
      $^{43}$Sc &  $-$6.67 &  25.06 &  24.77  &   0.29 \\
      $^{43}$Ti &  $-$6.67 &  17.24 &  17.12  &   0.12 \\
      $^{43}$V &  $-$2.55 &   4.72 &    5.05$^*$ &  $-$0.32 \\
      $^{44}$Ca &  $-$4.99 &  38.93 &  38.91  &   0.02 \\
      $^{44}$Sc &  $-$8.26 &  35.07 &  34.47  &   0.60 \\
      $^{44}$Ti & $-$13.88 &  33.06 &  33.42  &  $-$0.37 \\
      $^{44}$V &  $-$8.26 &  19.16 &   18.94$^*$ &   0.22 \\
      $^{44}$Cr &  $-$4.99 &   7.11 &   7.84$^*$ &  $-$0.74 \\
      $^{45}$Ca &  $-$4.61 &  46.73 &  46.32  &   0.41 \\
      $^{45}$Sc & $-$10.95 &  46.06 &  45.80  &   0.27 \\
      $^{45}$Ti & $-$15.49 &  43.09 &  42.95  &   0.14 \\
      $^{45}$V & $-$15.49 &  35.00 &   35.04  &  $-$0.04 \\
      $^{45}$Cr & $-$10.95 &  21.80 &  21.79$^*$ &   0.00 \\
      $^{45}$Mn &  $-$4.61 &   6.28 &   6.71$^*$ &  $-$0.43 \\
      $^{46}$Ca &  $-$6.73 &  56.90 &  56.72  &   0.18 \\
      $^{46}$Sc & $-$11.67 &  54.94 &  54.56  &   0.39 \\
      $^{46}$Ti & $-$20.14 &  56.02 &  56.14  &  $-$0.12 \\
      $^{46}$V$^{T=1}$ & $-$20.14 &  47.99 &   48.31  &  $-$0.32 \\
      $^{46}$V$^{T=0}$ & $-$19.77 &  47.75 &   47.51  &   0.24 \\
      $^{46}$Cr & $-$20.14 &  39.58 &  39.92  &  $-$0.35 \\
      $^{46}$Mn & $-$11.67 &  22.06 &  22.04$^*$ &   0.02 \\
      $^{46}$Fe &  $-$6.73 &   7.57 &   8.13$^*$ &  $-$0.56 \\
      $^{47}$Ca &  $-$6.10 &  64.20 &  63.99  &   0.21 \\
      $^{47}$Sc & $-$14.05 &  65.37 &  65.20  &   0.17 \\
      $^{47}$Ti & $-$21.06 &  65.11 &  65.02  &   0.09 \\
      $^{47}$V & $-$25.07 &  61.33 &   61.31  &   0.02 \\
      $^{47}$Cr & $-$25.07 &  52.98 &  53.08  &  $-$0.10 \\
      $^{47}$Mn & $-$21.06 &  40.05 &  40.00$^*$ &   0.05 \\
      $^{47}$Fe & $-$14.05 &  23.61 &  23.58$^*$ &   0.03 \\
      $^{48}$Ca &  $-$7.88 &  73.79 &  73.94  &  $-$0.15 \\
      $^{48}$Sc & $-$14.13 &  73.37 &  73.43  &  $-$0.07 \\
      $^{48}$Ti & $-$24.57 &  76.65 &  76.65  &   0.00 \\
      $^{48}$V & $-$27.58 &  71.99 &   71.85  &   0.14 \\
      $^{48}$Cr & $-$32.95 &  69.20 &  69.41  &  $-$0.21 \\
      $^{48}$Mn & $-$27.58 &  55.03 &  54.81$^*$ &   0.22 \\
      $^{48}$Fe & $-$24.57 &  42.72 &  43.14$^*$ &  $-$0.42 \\
      $^{48}$Co & $-$14.13 &  22.48 &  22.61$^*$ &  $-$0.13 \\
      $^{49}$Sc & $-$16.19 &  83.23 &  83.57  &  $-$0.34 \\
      $^{49}$Ti & $-$24.81 &  84.80 &  84.79  &   0.01 \\
      $^{49}$V & $-$31.01 &  83.45 &   83.40  &   0.04 \\
      $^{49}$Cr & $-$35.59 &  79.98 &  79.99  &  $-$0.01 \\
      $^{49}$Mn & $-$35.59 &  71.37 &  71.49  &  $-$0.13 \\
      $^{49}$Fe & $-$31.01 &  57.61 &  57.68$^*$ &  $-$0.07 \\
      $^{49}$Co & $-$24.81 &  41.74 &  41.90$^*$ &  $-$0.15 \\
      $^{50}$Ti & $-$27.72 &  95.49 &  95.73  &  $-$0.24 \\
      $^{50}$V & $-$32.16 &  92.50 &   92.74  &  $-$0.24 \\
      $^{50}$Cr & $-$40.54 &  92.95 &  92.99  &  $-$0.05 \\
      $^{50}$Mn$^{T=1}$ & $-$40.54 &  84.39 &  84.58  &  $-$0.18 \\
      $^{50}$Mn$^{T=0}$ & $-$40.28 &  84.26 &  84.35  &  $-$0.09 \\
      $^{50}$Fe & $-$40.54 &  75.47 &  75.64  &  $-$0.18 \\
      $^{50}$Co & $-$32.16 &  57.55 &  57.59$^*$ &  $-$0.04 \\
      $^{50}$Ni & $-$27.72 &  43.06 &  43.40$^*$ &  $-$0.34 \\
      $^{51}$V & $-$35.31 & 103.42 &  103.79  &  $-$0.37 \\
      $^{51}$Cr & $-$41.81 & 102.10 & 102.25  &  $-$0.15 \\
      $^{51}$Mn & $-$46.17 &  98.16 &  98.26  &  $-$0.11 \\
      $^{51}$Fe & $-$46.17 &  89.29 &  89.46  &  $-$0.17 \\
      $^{51}$Co & $-$41.81 &  75.51 &  75.74$^*$ &  $-$0.23 \\
      $^{51}$Ni & $-$35.31 &  59.09 &  59.12$^*$ &  $-$0.03 \\
      $^{52}$Cr & $-$45.99 & 114.05 & 114.29  &  $-$0.24 \\
      $^{52}$Fe & $-$54.27 & 105.46 & 105.64  &  $-$0.19 \\
      $^{52}$Ni & $-$45.99 &  78.08 &  78.41$^*$ &  $-$0.32 \\
      $^{54}$Fe & $-$62.85 & 129.63 & 129.71 &  $-$0.07 \\
      $^{54}$Co & $-$62.85 & 120.57 & 120.68 &  $-$0.11 \\
      $^{54}$Ni & $-$62.85 & 111.15 & 111.10 &   0.05 \\
      $^{56}$Ni & $-$78.46 & 142.44 & 141.94 &   0.50 \\
    \end{tabular}
  \end{center}
\end{table}

\begin{table}[h]
  \begin{center}
    \leavevmode
    \caption{Shell model binding energies relative to $^{40}$Ca
      compared with experiment. KB3 interaction with
      mass dependence, see text for the details}
    \begin{tabular}{ldddd}
      Nucleus & $E(SM)$ & $B_e$(The) & $B_e$(Exp) & $\Delta$ \\
      \hline
      $^{42}$Ca &  $-$2.71 &  19.83 &  19.84  &  $-$0.01 \\
      $^{42}$Sc$^{T=1}$ &  $-$2.71 &  12.34 &  12.64  &  $-$0.30 \\
      $^{42}$Sc$^{T=0}$ &  $-$2.35 &  12.21 &  12.02  &   0.19 \\
      $^{42}$Ti &  $-$2.71 &   4.45 &   4.85  &  $-$0.40 \\
      $^{43}$Ca &  $-$2.53 &  28.06 &  27.78  &   0.28 \\
      $^{43}$Sc &  $-$6.62 &  25.07 &  24.77  &   0.30 \\
      $^{43}$Ti &  $-$6.62 &  17.25 &  17.12  &   0.12 \\
      $^{43}$V &  $-$2.53 &   4.59 &    5.05$^*$ &  $-$0.45 \\
      $^{44}$Ca &  $-$4.92 &  38.76 &  38.91  &  $-$0.14 \\
      $^{44}$Sc &  $-$8.14 &  35.08 &  34.47  &   0.60 \\
      $^{44}$Ti & $-$13.67 &  33.08 &  33.42  &  $-$0.34 \\
      $^{44}$V &  $-$8.14 &  19.16 &   18.94$^*$ &   0.23 \\
      $^{44}$Cr &  $-$4.92 &   6.94 &   7.84$^*$ &  $-$0.90 \\
      $^{45}$Ca &  $-$4.51 &  46.58 &  46.32  &   0.25 \\
      $^{45}$Sc & $-$10.72 &  46.05 &  45.80  &   0.25 \\
      $^{45}$Ti & $-$15.16 &  43.13 &  42.95  &   0.18 \\
      $^{45}$V & $-$15.16 &  35.05 &  35.04   &   0.01 \\
      $^{45}$Cr & $-$10.72 &  21.78 &  21.79$^*$ &  $-$0.01 \\
      $^{45}$Mn &  $-$4.51 &   6.13 &   6.71$^*$ &  $-$0.58 \\
      $^{46}$Ca &  $-$6.54 &  56.73 &  56.72  &   0.02 \\
      $^{46}$Sc & $-$11.35 &  54.97 &  54.56  &   0.41 \\
      $^{46}$Ti & $-$19.57 &  56.01 &  56.14  &  $-$0.13 \\
      $^{46}$V$^{T=1}$ & $-$19.57 &  47.98 &  48.31   &  $-$0.33 \\
      $^{46}$V$^{T=0}$ & $-$19.21 &  47.85 &  47.51   &   0.35 \\
      $^{46}$Cr & $-$19.57 &  39.57 &  39.92  &  $-$0.36 \\
      $^{46}$Mn & $-$11.35 &  22.09 &  22.04$^*$ &   0.05 \\
      $^{46}$Fe &  $-$6.54 &   7.41 &   8.13$^*$ &  $-$0.73 \\
      $^{47}$Ca &  $-$5.90 &  64.12 &  63.99  &   0.13 \\
      $^{47}$Sc & $-$13.58 &  65.39 &  65.20  &   0.19 \\
      $^{47}$Ti & $-$20.36 &  65.16 &  65.02  &   0.14 \\
      $^{47}$V & $-$24.20 &  61.38 &  61.31   &   0.07 \\
      $^{47}$Cr & $-$24.20 &  53.03 &  53.08  &  $-$0.05 \\
      $^{47}$Mn & $-$20.36 &  40.10 &  40.00$^*$ &   0.10 \\
      $^{47}$Fe & $-$13.58 &  23.64 &  23.58$^*$ &   0.05 \\
      $^{48}$Ca &  $-$7.58 &  73.74 &  73.94  &  $-$0.20 \\
      $^{48}$Sc & $-$13.59 &  73.50 &  73.43  &   0.07 \\
      $^{48}$Ti & $-$23.60 &  76.66 &  76.65  &   0.02 \\
      $^{48}$V & $-$26.49 &  72.10 &  71.85   &   0.25 \\
      $^{48}$Cr & $-$31.61 &  69.16 &  69.41  &  $-$0.25 \\
      $^{48}$Mn & $-$26.49 &  55.14 &  54.81$^*$ &   0.33 \\
      $^{48}$Fe & $-$23.60 &  42.74 &  43.14$^*$ &  $-$0.40 \\
      $^{48}$Co & $-$13.59 &  22.61 &  22.61$^*$ &   0.00 \\
      $^{49}$Sc & $-$15.48 &  83.41 &  83.57  &  $-$0.16 \\
      $^{49}$Ti & $-$23.70 &  84.95 &  84.79  &   0.16 \\
      $^{49}$V & $-$29.61 &  83.56 &  83.40   &   0.16 \\
      $^{49}$Cr & $-$33.96 &  80.02 &  79.99  &   0.03 \\
      $^{49}$Mn & $-$33.96 &  71.41 &  71.49  &  $-$0.08 \\
      $^{49}$Fe & $-$29.61 &  57.73 &  57.68$^*$ &   0.05 \\
      $^{49}$Co & $-$23.70 &  41.89 &  41.90$^*$ &   0.00 \\
      $^{50}$Ti & $-$26.34 &  95.66 &  95.73  &  $-$0.07 \\
      $^{50}$V & $-$30.57 &  92.76 &  92.74   &   0.02 \\
      $^{50}$Cr & $-$38.47 &  92.94 &  92.99  &  $-$0.05 \\
      $^{50}$Mn$^{T=1}$ & $-$38.47 &  84.39 &  84.58  &  $-$0.19 \\
      $^{50}$Mn$^{T=0}$ & $-$38.25 &  84.39 &  84.35  &   0.04 \\
      $^{50}$Fe & $-$38.47 &  75.46 &  75.64  &  $-$0.18 \\
      $^{50}$Co & $-$30.57 &  57.80 &  57.59$^*$ &   0.22 \\
      $^{50}$Ni & $-$26.34 &  43.23 &  43.40$^*$ &  $-$0.17 \\
      $^{51}$V & $-$33.37 & 103.71 & 103.79   &  $-$0.07 \\
      $^{51}$Cr & $-$39.49 & 102.27 & 102.25  &   0.02 \\
      $^{51}$Mn & $-$43.60 &  98.22 &  98.26  &  $-$0.04 \\
      $^{51}$Fe & $-$43.60 &  89.35 &  89.46  &  $-$0.11 \\
      $^{51}$Co & $-$39.49 &  75.68 &  75.74$^*$ &  $-$0.06 \\
      $^{51}$Ni & $-$33.37 &  59.38 &  59.12$^*$ &   0.27 \\
      $^{52}$Cr & $-$43.21 & 114.21 & 114.29  &  $-$0.08 \\
      $^{52}$Fe & $-$50.95 & 105.38 & 105.64  &  $-$0.27 \\
      $^{52}$Ni & $-$43.21 &  78.24 &  78.41$^*$ &  $-$0.16 \\
      $^{54}$Fe & $-$58.46 & 129.70 & 129.71 &  $-$0.01 \\
      $^{54}$Co & $-$58.46 & 120.63 & 120.68 &  $-$0.05 \\
      $^{54}$Ni & $-$58.46 & 111.21 & 111.10 &   0.11 \\
      $^{56}$Ni & $-$72.31 & 142.38 & 141.94 &   0.44 \\
    \end{tabular}
  \end{center}
\end{table}

\end{multicols}
\end{document}